\documentclass[journal, twocolumns]{IEEEtran}
\usepackage[utf8]{inputenc}
\usepackage{graphicx}
\usepackage{color}
\usepackage{soul}
\usepackage[acronym,shortcuts]{glossaries}
\usepackage{balance}

\newacronym{sc}{SC}{Smart Contract}

\title{Extorsionware: Exploiting Smart Contract Vulnerabilities for Fun and Profit}
\author{Alessandro~Brighente, Mauro~Conti~\IEEEmembership{Fellow,~IEEE,} and Sathish~Kumar~\IEEEmembership{Senior Member,~IEEE,} 
\thanks{A. Brighente and M. Conti are with the Department
of Mathematics and HIT Research Center, University of Padova, via Trieste 63, Padova, Italy e-mail: firstname.lastname@unipd.it.}
\thanks{S. Kumar is with Cleveland State University, Electrical Engg. and Computer Science, Cleveland, OH, USA.}
}

\begin{document}

\maketitle
\begin{abstract}

\acp{sc} publicly deployed on blockchain have been shown to include multiple vulnerabilities, which can be maliciously exploited by users.

In this paper, we present extorsionware, a novel attack exploiting the public nature of vulnerable \acp{sc} to gain control over the victim's \ac{sc} assets. Thanks  to  the  control  gained  over  the  SC,  the  attacker obliges  the  victim  to  pay  a  price  to  re-gain  exclusive  control  of the SC. 
\end{abstract}
\glsresetall
\section{Introduction}
\IEEEPARstart{S}{mart} Contracts are the main enablers of new blockchain-based services and decentralized applications thanks to their self-verification, self-execution, and tamper resistance. Due to these features, they are used by multiple untrusted parties to stipulate agreements that are publicly verifiable. The impact of this technology is demonstrated by the number of actively deployed \acp{sc}, whose number increased by $100$ times in the last couple of years~\cite{coin}. When deployed in permission-less blockchains, \acp{sc}' code is publicly available. Therefore, anyone having access to the blockchain can investigate the \ac{sc}'s code. This allows external actors to gain insights on the procedures and operations that \acp{sc} execute, both in terms of amount of assets transferred and in terms of parties involved. Such information can be exploited by an attacker to select a suitable target to maximize the reward associated with a possible attack. In particular, loopholes or vulnerabilities leading to recurrent \ac{sc} behaviors, can be exploited by an attacker to steal assets from the account connected to the victim's \ac{sc}. 

The vulnerabilities associated with backdoors or non-restricted access to systems are very well known these days. In particular, during the last years, a large number of systems and companies were subject to \textit{ransomware} attacks~\cite{ransom}. In these attacks, the accessibility to systems' resources is compromised by an attacker gaining access to the network and encrypting all the accessible sensitive information. The attacker exploits the fact that attacks towards robust cryptographic protocols have no solution other than the brute-force exploration of all possible keys, which however requires an unfeasible amount of time to find the correct key. The victim has no choice other than paying the required ransom to the attacker to get the key and gaining back access to the encrypted sensitive information. The number of ransomware attacks and the requested ransom amount significantly increased in 2021, with a last significant example from the U.S. department of justice that paid $2.3$ Million Dollars in rasnom in June $2021$~\cite{doj}. Therefore, the presence of vulnerabilities might not only impact on the system's functionalities, but also provide attackers means for economic attacks.

In this work, we present the \textit{extorsionware}, a novel attack methodology inspired by ransomware attacks and targeting \acp{sc}. The public nature of smart contracts together with the lack of a fully secure \ac{sc} development framework, provide attackers with means for continuous exploitation of the identified vulnerability. Thanks to this, the attacker might be able to cause a continuous damage that allows for the request of a ransom to stop the exploitation. This concept is similar to a ransomware. However, instead of denying access to restricted data via encryption, the attacker is able to undermine the security of the \ac{sc} and its related operations. We discuss throughout the paper some of the vulnerabilities that can be exploited and the damage they may cause. We also explain why there is no simple workaround at the victim's side. Although several tools for designing secure \ac{sc} are available for developers~\cite{wan2021smart}, a recent study showed that the majority of the currently deployed \acp{sc} exhibit some vulnerabilities~\cite{perez2021smart}. Therefore, our framework represents a serious threat and shall not be overlooked.


\section{extorsionware Basic Blocks}
The extorsionware merges a technology, i.e., \ac{sc}, with an attack methodology, i.e., ransomware. In this section we review both the  technology and the attack methodology. In particular, we first provide an overview of the \ac{sc} technology, explaining how it works and how it can be used. Then, we review the ransomware attack. 

\subsection{Smart Contract}\label{sec:sc}
\acp{sc} are low-level programs stored and running on a blockchain. They are usually part of an event driven program, i.e., a program whose actions depend on the stimuli received by other on-chain or off-chain actors~\cite{zou2019smart}. \acp{sc} can execute different set of actions, such as transfer of funds, issuing tickets, and registering vehicles. Thanks to their self-execution design, \acp{sc} do not need human monitoring. Rather, they autonomously execute actions including, among the others, the creation of other contracts. The terms of agreement between the involved parties determine the set of actions and actors that determine the design and coding of the \ac{sc}. Compared to other types of contracts, \acp{sc} provide advantages such as the immutability of the information and trust enforcement among untrusted parties thanks to the use of the blockchain technology. In fact, in order to revert a transaction or alter the information provided or shared by a \ac{sc}, a malicious actor would need to revert the entire blockchain. This feature holds also for the \ac{sc} itself, as it cannot be modified unless altering the information stored in the blockchain. Notice however that, as for all other software, a certain degree of freedom for modification after deployment is needed to fix possible bugs and upgrade the \ac{sc}. \acp{sc} provide transparency to parties, as no third regulatory party is involved and the encrypted record of transactions is shared among all the participants. A further advantage of their decentralization is the time needed to execute transactions. In fact, there is no need to wait for the third party regulatory approval before execution, reducing hence the delays compared to other centralized systems.

\acp{sc} can actively issue transactions or make \textit{external calls} to other \acp{sc} to delegate the transaction execution~\cite{wang2018overview}. \ac{sc} can also perform \textit{internal calls}, i.e., calls generated within a contract that are not recorded on chain. \acp{sc} are also allowed to receive input parameters and values to use in the function execution. To execute a \ac{sc}, the sender needs to pay a fee associated to the contract's computational cost. In Ethereum-based \acp{sc}, the fee is associated to units of \textit{gas}, and this value is agreed upon before the instruction execution. The gas is used to reward the miner of the block containing the transaction, and any remaining gas is refunded to the issuer. To avoid loopholes where transaction are never terminated due to infinite cycles, each \ac{sc} defines a gas limit that can be used for contract execution. An out of gas exception is thrown whenever the gas limit is reached. 

\acp{sc}' code can be stored in plaintext in the blockchain~\cite{eth}. Therefore, \acp{sc} stored in permissionless blockchains are publicly available, and can be read by anyone accessing the network. A similar concept holds for permissioned blockchain, where access to code is provided upon authorization. However, in both cases, the set of actions and triggering conditions of the \ac{sc} can be accessed by simply accessing to the block where it is stored. This feature is exploited to provide transparency to the involved parties. However, as we will later discuss, this also allows attackers to identify and exploit \ac{sc} vulnerabilities. 

\subsection{Ransomware Attack}\label{sec:ransom}
\begin{figure}[!h]
    \centering
    \includegraphics[width=1.1\columnwidth]{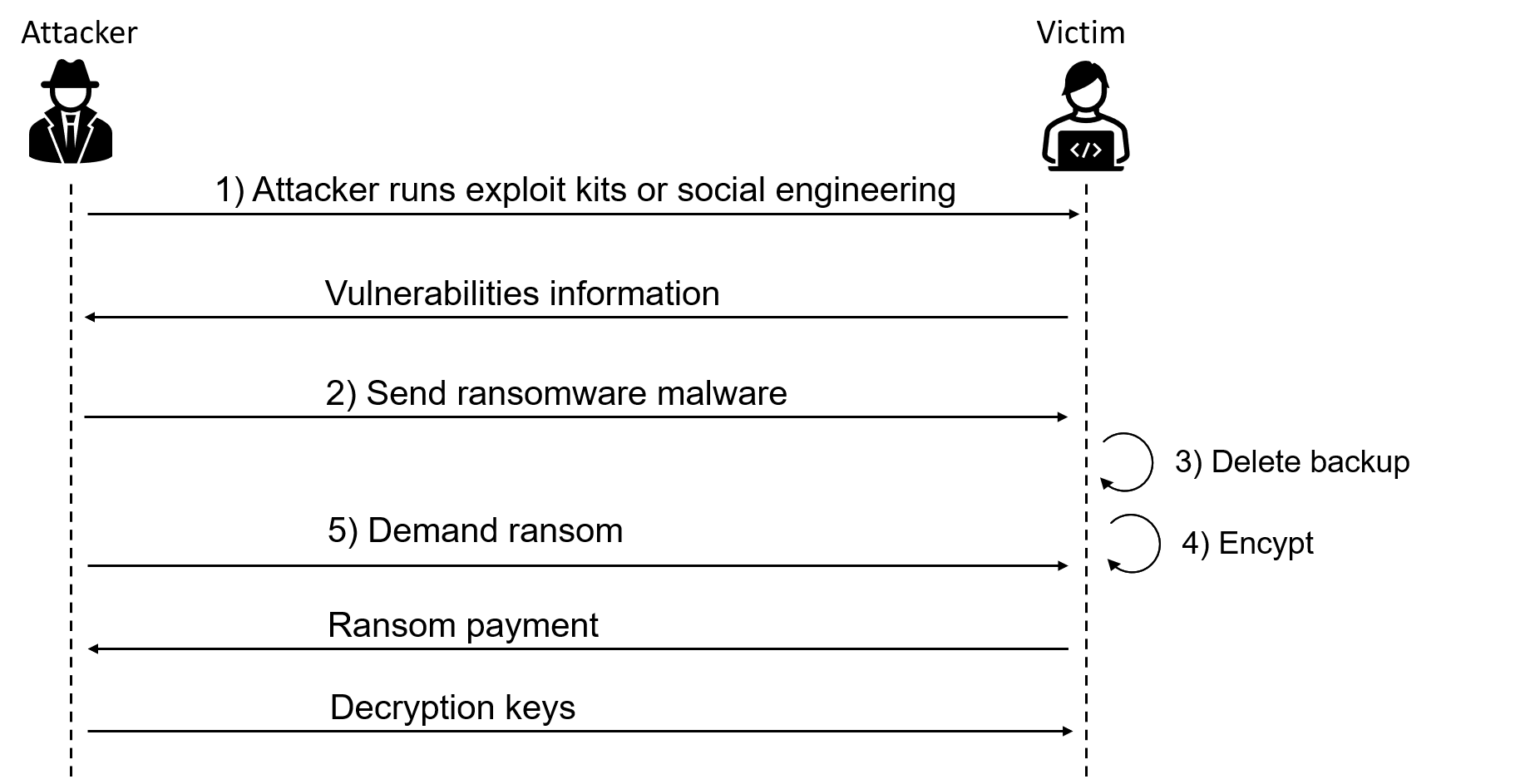}
    \caption{Five phases of a ransomware attack.}
    \label{fig:rnsm5phases}
\end{figure}

In a ransomware attack, a malicious user finds and exploit a vulnerability that allow accessing to resources of a victim's system. In particular, the attacker renders inaccessible such resources to the victim, and demands for a ransom to return the stolen assets. The ransomware attack can be divided in five phases~\cite{brewer2016ransomware}, as depicted in Figure~\ref{fig:rnsm5phases}.
The five phases can be described as follows:
\begin{enumerate}
    \item \textbf{Exploitation and Infection.} In this phase, the attacker either discovers a vulnerability in a target system, or injects malicious software in the victim's systems via social engineering attack (e.g., phishing). Exploitation can be performed also via exploit kit, which explore and identify common system's vulnerabilities. More uncommon (although likely) cases include leaking of the victim's sensitive information such as passwords for accessing the system.
    \item \textbf{Delivery and Execution.} In this phase, the attacker sends the actual ransomware executable and runs it on the victim's system. The attacker usually exploits strong encryption to prevent the identification of the executable. Most of these malware are equipped with persistence mechanisms, such that in case of reboot during the encryption process, the malware can pick up from where it has been interrupted. 
    \item \textbf{Back-up Spoliation.} This phase is peculiar of ransomware attacks. The malware, after few seconds from its execution, removes all back-up files or folders, as well as shadow copies to prevent the victim to recover the encrypted data without paying the ransom. The ransomware malware is usually able to remove eventual lockers on such files.
    \item \textbf{Encryption.} After deleting all back-up copies, the ransomware malware starts encrypting files on the victim's system. Depending on the malware implementation, the encryption keys are either agreed with the command and control server, or without reaching the internet. Encryption is performed via strong mechanisms, such as AES 256, such that the victim is not able to break the encryption without knowing the keys. Depending on the type of system and its connection with other network components, the malware may be able to encrypt files distributed over multiple network nodes.
    \item \textbf{User Notification and Clean-Up.} After successfully delivering the attack, the ransomware software usually presents to the victim instructions on how to recover files. The victim is given a couple of days, after which the ransom amount progressively increases. Finally, the malware removes itself from the system to avoid the collection of forensics information.
\end{enumerate}

\section{Smart Contract Vulnerabilities}
Similar to classical contracts written on paper, \acp{sc} need to be written by humans. The main difference is that, in the case of \ac{sc}, the writer is a programmer that translates the set of instructions and obligations stipulated by the involved parties into executable code. Statistics show that, due to human fallibility, codes contain a large number of bugs. An estimate reports an average number of $15$ - $50$ errors per $1000$ lines of delivered code \cite{mcconnell2004code}. Being \acp{sc} computer programs, they are hence susceptible to human mistakes in code design or writing. In this section we review some of the common coding mistakes resulting from the lack of expertise in the best \ac{sc} coding practices. All these vulnerabilities could be potentially exploited either singularly or simultaneously by attacker to perform extorsionware. The full list of known \ac{sc} vulnerabilities is provided by the Smart Contract Weakness registry~\cite{swc}. The gist of the SC vulnerabilities are described as follows.

\paragraph{Arithmetic Under/Overflow} The range of numbers that can be represented by a \ac{sc} depends on its code definition and is completely handled at compilation time. Therefore, in case of lack of a proper verification mechanism, an attacker can input data values that go beyond the range of the supported type.

\paragraph{Public Default Visibility} In Solidity, \acp{sc}' functions can be called by other contracts or users both internal and external to the blockchain depending on the visibility type. The default option is to set functions as public. Therefore, if visibility types are not properly handled, a malicious actor could call functions to trigger transactions execution.

\paragraph{Lack of True Randomness} The random number generation function is not present in solidity, as achieving decentralized randomness is a current research problem. Therefore, \ac{sc} programmers create their own random function. However, the lack of a properly design randomness source may lead to leakage of information that an attacker can exploit for predictions.  

\paragraph{Re-Entrancy} Ethereum \ac{sc} envision the possibility of calling and employing functionalities embedded in other \acp{sc}. A relevant example in this context, is the transfer of Ether from an address to another only via calls made to external \acp{sc}. A malicious user can exploit this feature to withdraw all the account's Ether. In particular, the attacker can create a malicious \ac{sc} that, upon receiving ETH, calls-back the victim's \ac{sc} to re-execute the transaction.

\paragraph{Locked Ether} This vulnerability exploits the fact that a \ac{sc} can be designed to execute fund transfers only by exploiting calls to other \acp{sc} similar to the previous point. However, in this case the attacker exploits a vulnerability in the executing \ac{sc} to delete it without permission~\cite{krupp2018teether}. The victim \ac{sc} therefore is not able to execute transactions as it makes calls to a non-existing \ac{sc}. Similar vulnerabilities that allow for the modification of data stored inside the \ac{sc} can be exploited to alter the address of the executing \ac{sc}. In this case, the attacker may change the address with that of a fictitious \ac{sc} that does not execute the transaction upon request from the victim. 

\paragraph{Unhandled Exceptions} This vulnerability is associated to pre-compiled \acp{sc}. Such contracts can return either zero or one as exit state, but may also report false-positives in case they generate an exception that is not properly handled by the \ac{sc}'s code.

\paragraph{Denial of Service} This vulnerability exploits external contract calls to stop the contracts' correct functioning. In particular, it exploits the fact that the victim's \ac{sc} may exploit multiple external calls but fails in properly handling exception during this operation. This might prevent the contract to continue its execution remaining in an idle state.

\section{extorsionware Life-Cycle}
The fact that \acp{sc}' code is publicly available provides malicious users unique opportunities to exploit possible vulnerabilities. In fact, it is reasonable to assume that a malicious users takes advantage of the identified vulnerability to attack the \ac{sc}. In this work, we present the extorsionware, a novel form of ransomware attack that targets the vulnerability of a publicly available \ac{sc}. Similarly to a ransomware, we assume that the attacker possesses an information that is essential to the victim, and that such information is returned to the victim only upon payment of a ransom. Differently from ransomware, we do not assume that the attacker is able to hide information at the victim's side (e.g., via encryption). We instead assume that the attacker identifies a vulnerability in a publicly available \ac{sc}, and exploits it and the fact that the victim is unaware of such vulnerability to request a ransom.

The extorsionware begins with an individual or company creating a \ac{sc} used for the transfer of funds and goods. The attack then includes the following five phases.
\begin{enumerate}
    \item \textbf{Scouting for Vulnerable SCs.} The attacker first crawls the blockchain to identify vulnerable \acp{sc} belonging to potential victims. As the \ac{sc} is publicly available, the attacker is able to identify possible vulnerabilities within the \ac{sc} code that allows her to transfer assets at will or to steal assets from the victim. 
    \item \textbf{Zero-Knowledge Attack.} In order for the attacker to show that she actually controls the victim's \ac{sc}, the ransom request comes with a certain operation executed from the victim's \ac{sc}. For instance, the attacker states that she will withdraw a specific amount from an account belonging to that \ac{sc} at a specific future time instant. Once the attacker executes the transaction at the specified day and time, the victim knows that the attacker controls the \ac{sc}. This scheme is similar to a zero-knowledge proof, where one of the parties discloses the knowledge of a certain information without actually disclosing the information itself.
    \item \textbf{Compensation Request and Ultimatum.} The attacker then gives the company an ultimatum for paying a ransom, or the vulnerability will be continuously exploited for malicious purposes or sold to the black market.
    \item \textbf{Continuous Exploitation.} The longer time goes on and the ransom remains unpaid, the amount transferred to others will continue to grow or the \ac{sc}'s functionalities will remain blocked. This is similar to how a typical ransomware model works in that the victim is motivated by a timer to quickly pay the ransom to prevent greater consequences.
    \item \textbf{Final Threat and Ransom Payment.}  Finally, the attacker will threaten to sell the \ac{sc} flaw to the black market if the ransom is never received. Once the ransom is paid, the attacker can reveal the flaw to the company. The company is then be able to rewrite the \ac{sc} code in order to stop future attacks from happening.
\end{enumerate}

\begin{figure}[!h]
    \centering
    \includegraphics[width=\columnwidth]{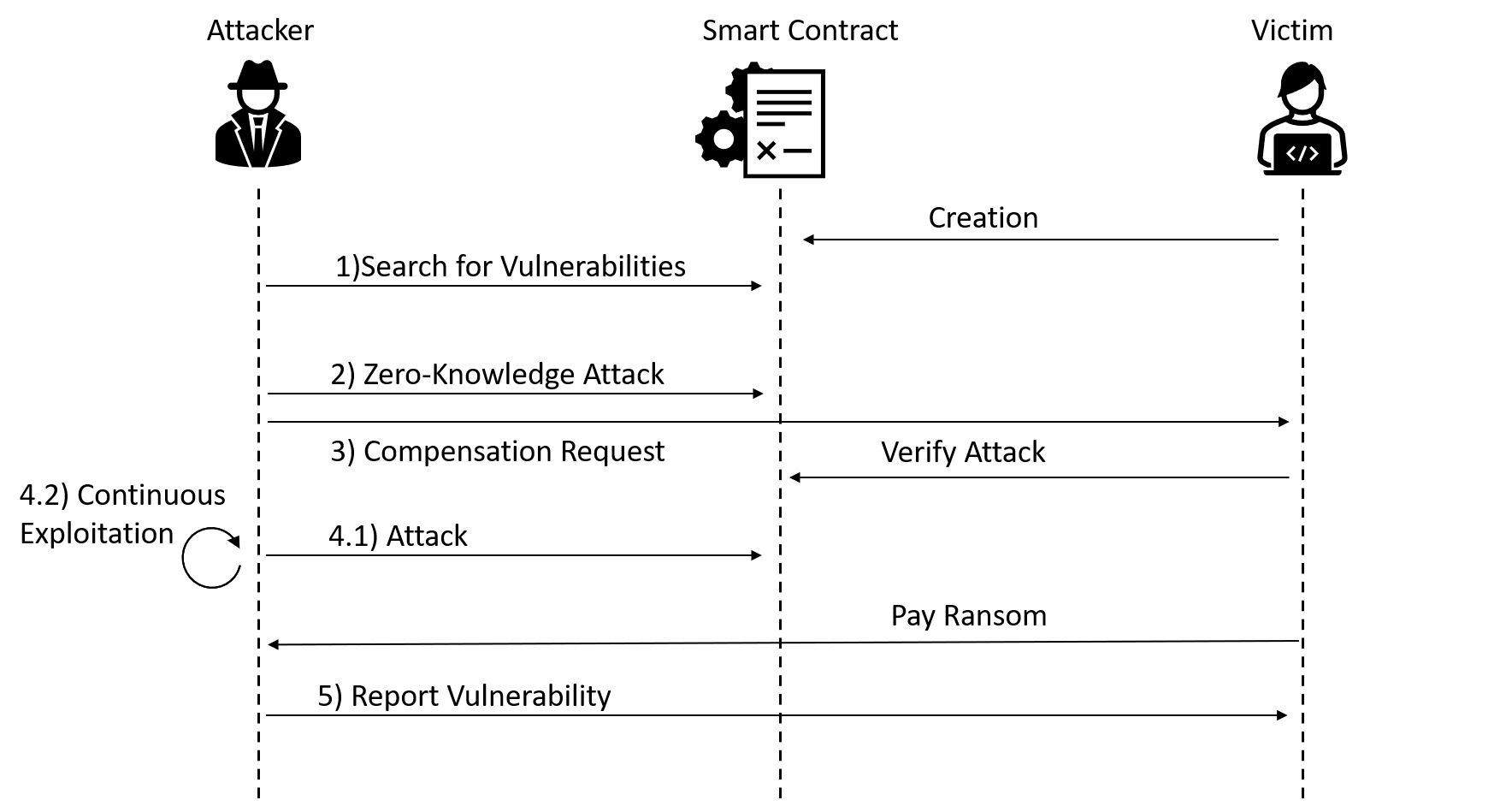}
    \caption{Life-cycle of the extorsionware attack. A victim unknowingly creates a vulnerable \ac{sc} and deploys it on the blockchain. An attacker accesses the code of the \ac{sc} and finds a vulnerability to exploit to demand for a ransom. The attacker exploits this vulnerability until the victim's payment has been received. }
    \label{fig:model}
\end{figure}

Figure~\ref{fig:model} shows the steps of extorsionware attack. Notice that the attack, depending on its purposes, is either continuous in time or is periodically launched until the ransom payment. Due to this, we divided the fourth step of extorsionware in $4.1)$ attack, and $4.2)$ continuous exploitation. 

An example of this attack would be an insurance company creating a \ac{sc} for paying customers. If an attacker found a vulnerability that allowed them to send money at will, they could create a devastating ransomware attack. The attacker could send \$1000 to another person as proof of this flaw and demand a ransom from the company of \$1 Million in order to reveal the flaw and stop sending out money. The insurance company may want to save its reputation early and pay the ransom. If the company does not think the financial loss of the \$1000 payouts is great enough yet, the attacker can increase the money sent to \$10,000 and keep increasing it until the ransom is paid. The attacker could also sell the flaw in the \ac{sc} on the black market so others could also take advantage of the vulnerability. We notice that a company may want to remove the flawed \ac{sc} and replace it with another one. However, deploying a new \ac{sc} comes with design and deployment costs. Furthermore, the legal implications of deploying a novel contract after failure of the originally stipulated one might not ease a \ac{sc} replacement. Lastly, depending on the service offered by the victim company, revealing that the contract has been attacked due to poor implementation might impact on the company's reputation. Therefore, it might be convenient for the victim to pay the ransom and recover the originally deployed contract. 

A further example on how an attacker can enforce the creation of weak \acp{sc} is provided by the large amount of information available online. It is indeed common that non-expert users are tricked into creating and deploying vulnerable \acp{sc} thanks to false promises of easy monetary reward. A further push through quick adoption of a technology without a proper understanding on its implications is the \textit{fear of missing out}, a form of social anxiety that pushes individuals to stay continually connected with what others are doing~\cite{hodkinson2019fear}. This might be the case of companies adopting blockchain and all related technology in a rush towards a market share, relying on internal non-expert staff, or to external experts that might be malicious

\section{Showcase Implementation of the extorsionware}
\begin{figure}[!h]
    \centering
    \includegraphics[width=0.9\columnwidth]{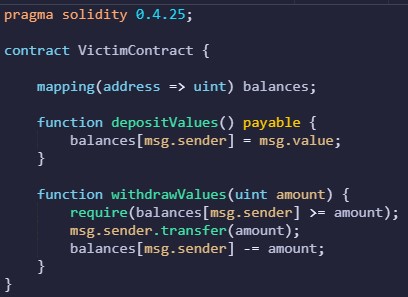}
    \caption{Solidity implementation of the victim's SC. The \ac{sc} requires external calls to perform transaction. However, it is vulnerable in terms of visibility and allows for re-entrance through the \textit{withdrawValues} function.}
    \label{fig:victim}
\end{figure}

In this section we show how one of the above mentioned vulnerabilities can be used for the extorsionware. Notice that our scope is not to showcase a new attack towards a specific \ac{sc} implementation. Rather, we use an already known vulnerability to demonstrate the validity of the extorsionware and shows how our framework can be exploited to demand a ransom. Furthermore, notice that an analysis of the vulnerabilities of currently publicly deployed \acp{sc} has already been provided~\cite{perez2021smart}. We exploit these results to validate the concreteness of the threat imposed by extorsionware.
We focus on the threat imposed by re-entrancy attacks together with public visibility. In particular, the attacker finds a vulnerability that allows to call the victim contract multiple times and withdraw a certain amount of Ether. We assume that the victim already agreed on the terms of the \ac{sc} with the other involved parties and deployed the \ac{sc} on the Ethereum blockchain\footnote{In this work we refer to the Ethereum blockchain as it is the most commonly used for SC deployment. However, our attack framework does not depend on the underlying blockchain implementation.}. 
Figure~\ref{fig:victim} shows the victim's Solidity\footnote{https://soliditylang.org/} \ac{sc} implementation. The victim's \ac{sc} is designed with a \textit{withdrawValues} function that allows for extracting a certain amount of Ether from the user's account after calling an external contract via the \textit{transferValues} function. Notice that no visibility has been specified for this function, therefore it assumes the default public visibility and can be controlled by external contracts. The vulnerability in this contract resides in the fact that the transfer function hands over the control to the receiving \ac{sc}. If the latter is controlled by a malicious user, it could call back the sender before the transaction is completed.

\begin{figure}[!h]
    \centering
    \includegraphics[width=0.8\columnwidth]{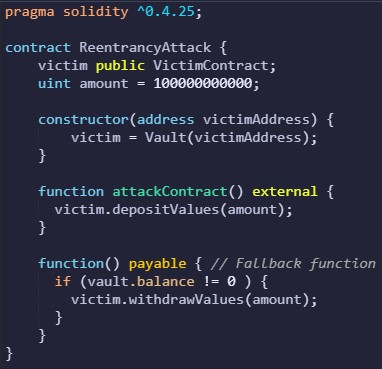}
    \caption{Solidity implementation of the attacker's SC. The attack exploits the re-entrance vulnerability to require multiple transfers from the victim's \ac{sc} though the function \textit{payable}.}
    \label{fig:attacker}
\end{figure}

Figure~\ref{fig:attacker} shows the Solidity implementation of the attacker's \ac{sc}. This \ac{sc} exploit the aforementioned vulnerability, i.e., the call back to the sender's \ac{sc}. The attack starts with a call to the victim's withdraw function. When the victim's \ac{sc} call the transfer function, this will trigger in any receiving contract a \textit{payable} fallback. In the contract in Figure \ref{fig:attacker}, the payable function calls back the victim's withdraw function. The victim hence repeatedly triggers call to the attacker's contract until all Ethers are drawned. In extorsionware, the attacker may stop this loop to prevent all assets to be withdrawn, or may start the process again if the ransom is not paid. This represents a form of zero-knowledge proof, where the attacker proves the knowledge of a secret to the victim without sharing the secret. We stress the fact that, although the ransom may imply a monetary loss higher than the amount of Ether withdrawn by the \ac{sc}, a company may still want to pay the ransom to avoid a costumer being deprived of her or his assets with a consequent loss of reputation by the company itself. 

Upon reception of the ransom payment, the attacker shares with the victim the information on the exploited vulnerabilities. The \ac{sc} is then upgraded to fix the bugs and prevent further exploitation of the same vulnerability by other malicious actors. Notice that, although from the perspective of the vulnerabilities exploited in this showcase implementation, it might have been rather simple for an expert programmer to identify the problem and correct the vulnerabilities without paying a ransom. However, this might not always be the case in larger and more complex \acp{sc}, where the vulnerability might be hidden among a large number of lines of code. 
\section{Conclusions and Future Works}
Although the public availability of \acp{sc}' code enforces trust among non-trusted parties, it also opens up possibilities for attackers to exploit potential \ac{sc} implementation weaknesses. In this work we presented the extorsionware, a novel attack framework inspired by ransomware attacks that exploits \ac{sc} vulnerabilities to demand for ransoms. With our framework, an attacker is able to both show the knowledge of a weakness by programming attacks with specific effects, and to enforce victims to pay ransoms instead of removing publicly deployed contract. Due to the legal implications of \acp{sc}, a victim could not simply remove the \ac{sc} and create a new one without incurring in economic or credibility losses. We belive that the threat imposed by extorsionware should be considered to enforce the design of secure \acp{sc} to avoid companies and privates exploiting this technology incurring in uncomfortable and risky situations. 

In future works we will assess the impact that different types of vulnerability have when exploited by extorsionware. In particular, we will focus on the controllability provided to the attacker. Furthermore, we plan to develop a countermeasure for extorsionware. An idea might be to build a shield \ac{sc} able to detect whether an exploitation is occurring. The shield \ac{sc} will also serve to block the exploitation point and exclude the attacker from the network.

\balance
\bibliographystyle{IEEEtran}
\bibliography{bib}

\begin{IEEEbiographynophoto} {A. Brighente} is a postdoctoral researcher of the Security and Privacy research group (SPRITZ) at the University of Padova.  He received his Ph.D. degree in Information Engineering from the University of Padova in Feb. 2021. From June to November 2019 he was visitor researcher at Nokia Bell Labs, Stuttgart. He published several papers in top-most conferences and journals, and he is also author of a patent. He has been involved in European projects (Ontochain) and company projects (IOTA) with the University of Padova. He served as TPC for several conferences, including Globecom and CANS. He is guest editor for IEEE Transactions on Industrial Informatics. His current research interests include security and privacy in cyber-physical systems, UAV communications, vehicular networks, blockchain/distributed ledger technology, and physical layer security.
\end{IEEEbiographynophoto}

\begin{IEEEbiographynophoto}{ M. Conti} is Full Professor at the University of Padua, Italy. He is also affiliated with TU Delft and University of Washington, Seattle. He obtained his Ph.D. from Sapienza University of Rome, Italy, in 2009. After his Ph.D., he was a Post-Doc Researcher at Vrije Universiteit Amsterdam, The Netherlands. In 2011 he joined as Assistant Professor the University of Padua, where he became Associate Professor in 2015, and Full Professor in 2018. He has been Visiting Researcher at GMU, UCLA, UCI, TU Darmstadt, UF, and FIU. He has been awarded with a Marie Curie Fellowship (2012) by the European Commission, and with a Fellowship by the German DAAD (2013). His research is also funded by companies, including Cisco, Intel, and Huawei. His main research interest is in the area of Security and Privacy. In this area, he published more than 400 papers in topmost international peer-reviewed journals and conferences. He is Area Editor-in-Chief for IEEE Communications Surveys \& Tutorials, and has been Associate Editor for several journals, including IEEE Communications Surveys \& Tutorials, IEEE Transactions on Dependable and Secure Computing, IEEE Transactions on Information Forensics and Security, and IEEE Transactions on Network and Service Management. He was Program Chair for TRUST 2015, ICISS 2016, WiSec 2017, ACNS 2020, and General Chair for SecureComm 2012, SACMAT 2013, CANS 2021, and ACNS 2022. He is Senior Member of the IEEE and ACM. He is a member of the Blockchain Expert Panel of the Italian Government. He is Fellow of the Young Academy of Europe.
\end{IEEEbiographynophoto}

\begin{IEEEbiographynophoto}{S. A. P. Kumar}  is an Associate Professor of Computer Science in the Department of Electrical Engineering and Computer Science at the Cleveland State University, Cleveland, Ohio, USA. He is co-director of Cleveland State University TECH Hub, an interdisciplinary center related to technology. He is directing Intelligent Secure Cyber-Systems Analytics and Applications Research (ISCAR) Lab at the Cleveland State University. He earned his PhD degree in Computer Science and Engineering from the University of Louisville, Kentucky, USA in 2007. He is a Senior Member of IEEE. His current research interests are in cybersecurity, machine learning, big data analytics and secure distributed systems and their applications. He has published more than 50 technical papers in international journals and conference proceedings.  
\end{IEEEbiographynophoto}

\end{document}